\documentclass[runningheads]{llncs}

\usepackage[ruled,vlined]{algorithm2e}
\usepackage{tikz}
\usepackage{etoolbox}

\usepackage{amssymb,amsmath} 
\usepackage[strings]{underscore} 
\usepackage{subcaption} 
\usepackage{comment}

\AtBeginDocument{%
  \providecommand\BibTeX{{%
    \normalfont B\kern-0.5em{\scshape i\kern-0.25em b}\kern-0.8em\TeX}}}

\newcommand{\mC}{\mathcal{C}} 
\newcommand{\mX}{\mathcal{X}} 
\newcommand{\mI}{\mathcal{I}} 
\newcommand{\R}{\mathbb{R}} 
\newcommand{\xb}{\mathbf{x}} 
\newcommand{\yb}{\mathbf{y}} 
\newcommand{\zb}{\mathbf{z}} 
\newcommand{\mG}{\mathcal{G}} 
\newcommand{\mN}{\mathcal{N}} 
\newcommand{\mS}{\mathcal{S}} 
\newcommand{\nmin}{N_{\min}}
\newcommand{\nmax}{N_{\max}}
\newcommand{\smin}{\mS_{\min}}
\newcommand{\smax}{\mS_{\max}}

\newcommand{\mCm}{\mC_{k_1}, \ldots, \mC_{k_m}} 
\newcommand{\mCi}[1]{\mC_{k_#1}} 
\newcommand{\xyi}[1]{\{(\xb^{(#1)}_1, y^{(#1)}_1), \ldots, (\xb^{(#1)}_{n_#1}, y^{(#1)}_{n_#1})\}} 
\newcommand{\xyistar}[1]{(\xb_*^{(#1)}, y_*^{(#1)})} 
\newcommand{\fyi}[1]{f_y^{(#1)}} 
\newcommand{\gyi}[1]{g_y^{(#1)}} 
\newcommand{\mVi}[1]{\mathcal{V}_{(#1)}} 

\newbool{arxiv}
\booltrue{arxiv}

\begin{document}

\title{Scalable Statistical Inference of Photometric Redshift via Data Subsampling}

%
\author{Arindam Fadikar\inst{1}
\orcidID{0000-0001-7396-0350} \and
Stefan M.\ Wild\inst{1}
\orcidID{0000-0002-6099-2772} \and Jonas Chaves-Montero\inst{2} \orcidID{0000-0002-9553-4261}}
\authorrunning{A. Fadikar et al.}
%
\institute{$^1$Mathematics and Computer Science Division, 
Argonne National Laboratory, Lemont IL 60439, USA\\
\email{\{afadikar,wild\}@anl.gov}\\
$^2$Donostia International Physics Centre, Paseo Manuel de Lardizabal 4, 20018
Donostia-San Sebastian, Spain.\\
\email{jonas.chaves@dipc.org}}
\maketitle             

\begin{abstract}
  Handling big data has largely been a major bottleneck in traditional statistical models. Consequently, when accurate point prediction is the primary target, machine learning models are often preferred over their statistical counterparts for bigger problems 
But full probabilistic statistical models often outperform other models 
in quantifying uncertainties associated with model predictions. We develop a data-driven statistical modeling framework that combines the uncertainties from an ensemble of statistical models learned on smaller subsets of data carefully chosen to account for imbalances in the input space. We demonstrate this method on a photometric redshift estimation problem in cosmology, which seeks to infer a distribution of the redshift---the stretching effect in observing the light of far-away galaxies---given multivariate color information observed for an object in the sky. 
Our proposed method performs balanced partitioning, graph-based data subsampling across the partitions, and training of an ensemble of Gaussian process models.
\end{abstract}

\section{Introduction}

Data analysis techniques have become an essential part of a scientist's toolbox for making inferences about an  underlying system or phenomenon. With the advancement of modern computing, processing power has gone up manyfold, and data volumes have grown at a similar pace. Thus, there is a growing demand for scalable machine learning (ML) and statistical techniques that can handle large data in an efficient and reasonable way. Recent trends include the use of ML and statistical regression techniques such as deep neural networks~\cite{lawrence1997face,hu2002handbook}, tree-based models~\cite{hastie2009random}, and Gaussian process (GP) models~\cite{Rasmussen2005,gramacy2015local}. Each technique has advantages and disadvantages that promote or limit its use for a specific application. For example, deep neural networks are known for achieving superior prediction accuracy but at the cost of significant data and compute cost for training. 
Statistical models such as GPs approximate the global input-output relationship and provide full uncertainty estimates at a fraction of  the data required by deep neural networks.
However, statistical models do not generally scale well with data size. Some GP models have been proposed to find workarounds such as introducing sparse approximation of large correlation matrices~\cite{kaufman2011efficient}, using locally learned models~\cite{gramacy2015local} or considering only a subset of the data \cite{banerjee2012bayesian}.

In this paper we propose a statistical modeling framework that can handle large training data by leveraging data subsampling combined with advanced statistical regression techniques to provide a full density estimate. The basic idea entails using smaller subsets of data to train regression models and building an ensemble of these models. Our modeling approach differs from other approximations in that we attempt to learn the global input-output relationship in each individual subsample and model. This is counterintuitive and opposite  approaches that seek to build accurate local response surfaces~\cite{gramacy2015local,kaufman2011efficient}. However, uncertainty estimates can be undesirably constricted in locally learned models. Furthermore, data partitioning and subsampling in our proposed approach are driven entirely  by data distribution and are free from model influences present in other locally learned models. We focus on the analysis paradigm in cosmology that deals with estimation of the redshifts of objects (e.g., galaxies) as a motivating application, which is discussed in the following section.

The rest of the paper is structured as follows. In \S\ref{sec:case} we  describe the estimation problem and data to be used. Our proposed methodology is described in \S\ref{sec:method}, and numerical results are presented in \S\ref{sec:results}. In \S\ref{sec:discussion} we summarize our approach and its benefits.

\section{Photometric redshift estimation}
\label{sec:case}

The cosmological analysis of galaxy surveys, from gathering information on dark energy to unveiling the nature of dark matter, relies on the precise projection of galaxies from two-dimensional sky maps into the three-dimensional space~\cite{weinberg2013_ObservationalProbescosmica,kaufman2011efficient}. However, measuring the distance of distant objects in the Universe from the Earth is challenging. Furthermore, the accelerated expansion of the Universe causes the wavelength of light from a distant object to be stretched or \emph{redshifted}.
Interestingly, the redshift of an object is proportional to its receding speed and can be used to estimate the radial distance to this source. For an accurate redshift estimation, one would have to obtain the full spectrum of each galaxy at a very high resolution, which is a demanding and time-consuming task~\cite{ilbert2009_COSMOSPhotometricRedshifts,beck2016photometric}. An alternative to this approach is to infer the redshift based on the intensity
of galaxy light observed through a reduced number of wavebands~\cite{baum1957_PhotoelectricDeterminationsredshifts}. Such an approximation is known as photometric redshift, whereas the redshift obtained from the full spectrum is called spectroscopic redshift.

Photometric redshift estimation methods can be divided into two categories: spectral energy distribution template fitting~\cite{puschell1982_NearinfraredPhotometrydistant,Fernandezsoto:99} and statistical regression or ML methods~\cite{Firth:03,cavuoti2012_PhotometricRedshiftsquasi,kaufman2011efficient}. Supervised ML methods such as deep neural networks have recently seen success in approximating the mapping between broadband fluxes and redshift~\cite{beck2016photometric}. Many of the successes of ML methods, however, rely on the availability of large training datasets coupled with large computing power. Because of their increased computational complexity, statistical regression methods have not been a preferred choice, even when the data size is moderately large.

The case study 
we consider is the publicly available data release 7 of the Sloan Digital Sky Survey (SDSS)~\cite{York:00}, which contains approximately 1 million galaxies with spectroscopic redshifts estimates. 
From this initial sample, we select $\simeq100\,000$ galaxies with signal-to-noise ratio larger than 10 in all photometric bands, clean photometry, and high-confident spectroscopic redshift below $z=0.3$. The photometry of each SDSS galaxy consists of flux measurement in five broadband filters $u, g, r, i$, and $z$, which serve as input to the predictive model for redshift.

In general we consider a dataset of $N (=10^5)$ scattered data points $\xb_1, \ldots, \xb_N$ contained in a compact input space $\mX\subset \R^d$, with corresponding logged redshifts $y_1, \ldots, y_N$. Without loss of generality, we assume that the input space has been normalized so that $\mX=[0,1]^d$ is the unit cube and the logged redshift values are transformed to mean zero and a standard deviation of 1. In our case study, $d$ is 5, with the dimensions corresponding to four colors computed by taking the ratio of the flux consecutive filters and the magnitude in the i-band. In what follows, we refer to these variables as color space.

Approximating the relationship between broadband filter values and the photometric redshift is challenging because of several factors. For example, lack of coverage of the input space in the training dataset results in poor predictions at the extremes, as well as degeneracies arising in the prediction of the redshift in some cases. Hence, simple Gaussian predictive uncertainty may not accurately represent the distribution of the redshift given a set of colors. For example, a galaxy at a high redshift with high luminosity and a galaxy at a low redshift with low luminosity may register the same flux values to a telescope. Considering such uncertainty in the predictive distribution would require discovering and modeling these latent processes. Mixed-density networks~\cite{d2018photometric} and full Bayesian photometric redshift estimation~\cite{benitez2000bayesian} are  examples that model the predictive distribution by a finite number of Gaussians. While our objective is similar to~\cite{d2018photometric}, we give greater importance to obtaining a full predictive distribution that can accurately reflect multiple predictive modes.

\section{Statistical methodology}
\label{sec:method}

Our proposed approach consists of three intermediate stages: 
(1) partitioning the input color space $\mX$ into a user-defined number of partitions in order to ensure balance among the input data, (2) subsampling datapoints from these partitions, and (3) training a regression model on the sampled data. The final predictive model is then based on an ensemble of models obtained by repeated execution of the latter two stages. Although we present the basic implementations here for simplicity, scalability is emphasized in each stage: the partitioning can exploit domain decomposition parallelism, the subsampling allows for model construction using datasets of user-specified greatly reduced size, and the ensemble is naturally parallelizable.  

\subsection{Partitioning the input space}
\label{sec:partition}
We begin by partitioning the color space $\mX$ into a set of mutually exclusive subsets $\mC=\{\mC_k: k \in \mI\}$ such that $\cup_{k \in \mI} \mC_k = \mX$ and $\mC_{k_i} \cap \mC_{k_j} = \emptyset$ for $k_i\neq k_j$.

We especially target datasets that are highly nonuniform, including those arising in redshift estimation, and thus we seek balanced partitions. By balanced, here we intend for the number of training points in all partitions to satisfy $|\mC_k| \approx \frac{N}{|\mI|}$, where $|\mC_k|$ denotes the cardinality of $\mC_k$.

The number of partitions ($m$) is not predetermined for our balanced hyperrectangle partitioning algorithm. However, the values of the inputs to the algorithm, $\nmin$ and $\nmax$---minimum and maximum number of datapoints in each partition, respectively---are influenced by a desired $m$. To achieve a (n approximately) balanced hyperrectangle partition, we propose a three-step procedure that consists of an initialization of coarse partitions, pruning to satisfy the minimality condition  $\min_{k \in \mI} |\mC_k| \geq \nmin$, and then splitting to satisfy the maximality condition  $\max_{k \in \mI} |\mC_k| \leq \nmax$. We also require $\nmax \geq 2 \nmin$ to guarantee termination of our procedure.

\paragraph{Initialize} Interval boundaries $0=x^p_0 < x^p_1 < \ldots < x^p_{m_p}=1$ of size $m_p$ are defined along each input dimension $p =1, \ldots, d$. Then an initial set of hyperrectangle partitions $\mC$ in the $d$-dimensional input space is constructed as the Cartesian product of all $1D$ intervals:
\begin{align}
\label{eq:hyperrec-init}
 \mC = \left\{A_{j_1}^1 \times \cdots \times A_{j_d}^d :  j_p = 1, \ldots, m_p,\, p = 1, \ldots, d\right\}.   
\end{align}
This leaves open a choice of interval boundary values along each dimension. In this study, we have opted for quantile-based splits. The result of the initialization step is $\prod_{p=1}^d m_p$ hyperrectangles, some of which may be empty. 

\paragraph{Merge} In the next step, partitions with cardinality less than $\nmin$ are merged successively with their neighbors until the minimality condition is satisfied. Note that at the end of the merge step, some partitions may have cardinality greater than $\nmax$. The merging algorithm is briefly described below.

We define $\smin = \{ \mC_k \in \mC: |\mC_k| < \nmin \}$ to be the set of partitions with cardinality less than $\nmin$. We begin by identifying a target partition $\mC_{(0)} \in \arg \min_{\mC_k \in \smin} |\mC_k|$, namely, a partition with the smallest cardinality. We also define the directional neighborhood function $\mN^p_{\omega}(\cdot)$, which represents neighbors of $\cdot$ along dimension $p \in \{1, \ldots, d\}$ and where $\omega \in \{\text{lower} \equiv l, \text{upper} \equiv u\}$ denotes the relative position of the neighbors with respect to dimension $p$. 

Given $\mC_{(0)}$, the selection of partition(s) to merge with is equivalent to finding an appropriate dimension and direction to merge along. At most $2d$ such merging choices exist. We require that the newly formed partition be a hyperrectangle. For a given partition $\mC_{(0)}$, the merging dimension and direction are selected according to following condition:
\begin{align}
\label{eq:merge}
(p^*, \omega^*) \in \arg \min_{\substack{p = 1, \ldots, d \\ \omega \in \{l, u\}}} \sum_{\mC_k \in \mN^p_\omega(\mC_{(0)})} |\mC_k|,
\end{align}
in other words, the dimension and direction for which the updated partition contains the least number of datapoints among all possible combinations. In the case of multiple $(p^*, \omega^*)$ possibilities in \eqref{eq:merge}, the one that results in the most uniform sides after merging is selected. In particular, we choose the merger that results in the smallest ratio between the longest and shortest sides. Once the merged hyperrectangle achieves at least cardinality $\nmin$, the set $\smin$ is updated. The merging step continues until $\smin$ is empty.

\paragraph{Split} In the split step, partitions with cardinality greater than $\nmax$ are successively broken into smaller partitions until the maximality condition is satisfied. We define $\smax = \{ \mC_k : |\mC_k| > \nmax \}$ to be the set of hyperrectangles with cardinality greater than $\nmax$. A partition with highest cardinality in $\smax$ is selected for a split, $\mC_{(m)} \in \arg \max_{\mC_k \in \smax} |\mC_k|$. A split of a hyperrectangle is defined as breaking it into two hyperrectangles along one dimension. Hence, for any split operation the appropriate dimension needs to be identified. The break point in that dimension can be any location for which the two new partitions satisfy the minimality condition (such a location exists because we require $\nmax \geq 2 \nmin$). Our case study uses the median as the break point. To promote uniformity in the shape of hyperrectangle partitions, we always select the dimension corresponding to the longest side to perform the split, unless any of the resulting hyperrectangles breaks the minimality conditions, in which case we move to the next best dimension. Successive splits are carried out until $\smax$ is empty. 
\ifbool{arxiv}{
Algorithm~\ref{algo:hyperrec} summarizes our proposed method for hyperrectangle partitioning.

\begin{algorithm}[t]
    \SetAlgoLined
    \DontPrintSemicolon
    \SetKwInOut{Input}{input}
    \SetKwInOut{Output}{output}
    \SetKwInOut{Init}{initialize}
    \SetKwInOut{Merge}{merge}
    \SetKwInOut{Split}{split}
    \SetKwBlock{Begin}{begin}{end}
    \KwData{$(\xb_1, y_1), \ldots, (\xb_N, y_N)$}
    \Input{$\nmin$, $\nmax$}
    \Output{partitioning 
            $\mC$, induced graph $\mG$} 
    \vspace{0.2cm}
    {\Large \textbf{initialize}} \;
    \For{$p = 1, \ldots, d$}{
        \tcp{Define $m_p$ intervals along dimension $p$}
        $A_j^{p} = [x^p_{j-1}, x^P_j), \; j = 1, \ldots, m_p$ \;
    }
    \tcp{Define hyperrectangle partitions}
    $\mC = \left\{A_{j_1}^1 \times \cdots \times A_{j_d}^d :  j_p = 1, \ldots, m_p,\, p = 1, \ldots, d\right\}$ \;
    \vspace{0.2cm}
    {\Large \textbf{merge}} \;
    $\smin = \{ \mC_k \in \mC: |\mC_k| < \nmin \}$ \;
    \While{$\smin \neq \emptyset$}{
        \tcp{Find partitions to merge}
        $\mC_{(0)} \in \arg \min_{\mC_k \in \smin} |\mC_k|$ \;
        \tcp{Define new partitions and update $\smin$}
        \While{$|\mC_{(0)}| < \nmin$}{
            $(p^*, \omega^*) \in \arg \min_{\substack{p = 1, \ldots, d \\ \omega \in \{l, u\}}} \sum_{\mC_k \in \mN^p_\omega(\mC_{(0)})} |\mC_k|$ \;
            $\mC \gets \mC \backslash (\mC_{(0)} \cup \mN^{p^*}_{\omega^*}(\mC_{(0)}))$\;
            $\mC_{(0)} \gets \mC_{(0)} \cup \mN^{p^*}_{\omega^*}(\mC_{(0)})$\;
            $\mC \gets \mC \cup \mC_{(0)}$
        }
       $\smin = \{ \mC_k \in \mC: |\mC_k| < \nmin \}$ \; 
    }
    \vspace{0.2cm}
    {\Large \textbf{split}} \;
    $\smax = \{ \mC_k \in \mC: |\mC_k| > \nmax \}$ \;
    \While{$\smax \neq \emptyset$}{
        \tcp{Find partition and edge to split}
        $\mC_{k^*} \in \arg \max_{\mC_k \in \smax} |\mC_k| 
        \hfill =: A_{j_1}^1 \times \cdots \times A_{j_d}^d$ \;
        $p^* \in \arg \max_{p = 1, \ldots, d} \text{ length}(A^p_{j_p})$ \;
        $A^{p^*}_{j_{P^*, l}} \cup A^{p^*}_{j_{P^*, u}} = A^{p^*}_{j_{P^*}}$ \;
        \tcp{Define new partitions and update $\smax$}        
        $\mC_{(k^*,\omega)} = A_{j_1}^1 \times \cdots\times A^{p^*}_{j_{P^*, \omega}} \times \cdots \times A_{j_d}^d, \; \omega \in \{l,u\}$ \;
        $\mC \gets \mC \backslash \mC_{k^*} \cup \mC_{(k^*,l)} \cup \mC_{(k^*,u)}$\;
        $\smax=\{ \mC_k \in \mC : |\mC_k| > \nmax \}$
    }
 
 \caption{Balanced hyperrectangle partitioning}
 \label{algo:hyperrec}
\end{algorithm}
}{}

\begin{figure}
    \centering
    \includegraphics[width = \textwidth]{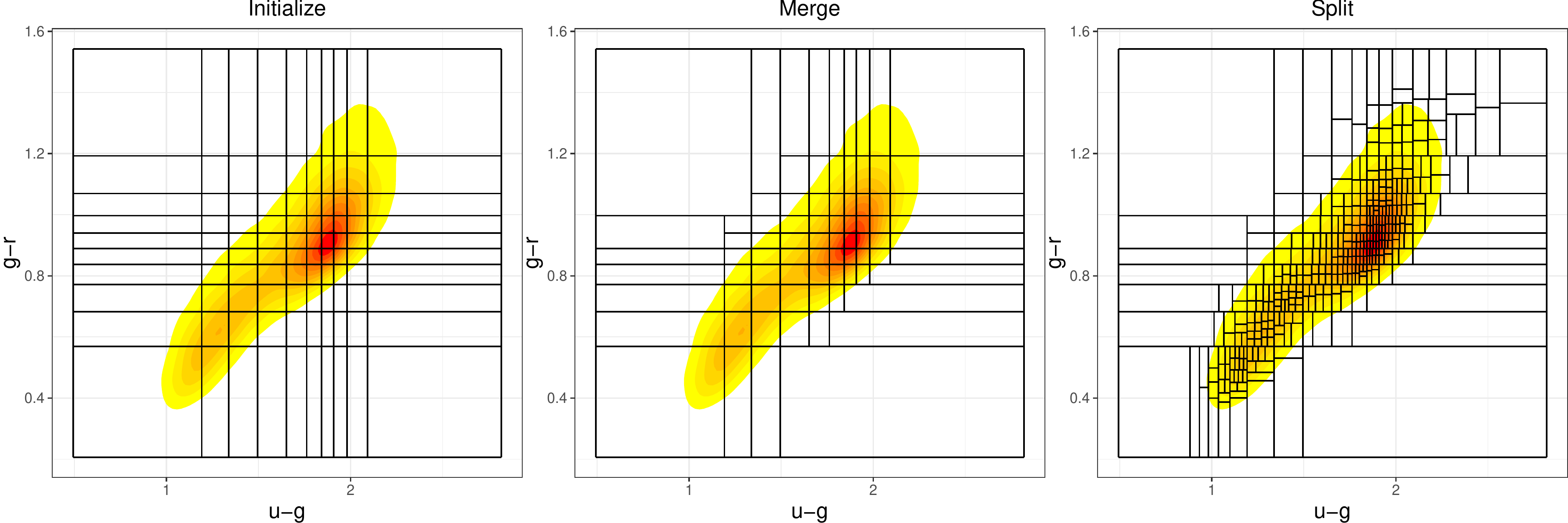} 
    \caption{Illustration of hyperrectangle partitioning of a $2D$ space containing approximately 80,000 datapoints, with input $(\nmin, \nmax) = (100,\, 300)$. At the beginning, 100 partitions are initialized based on marginal percentiles $10\%, \cdots, 90\%$ (left); merging then occurs for partitions with cardinality less than $\nmin$ (middle); and lastly partitions with cardinality more than $\nmax$ are successively split into smaller partitions, resulting in 393 partitions (right).}
    \label{fig:hyperrec-2d}
\end{figure}

\subsection{Conditional sampling from partitions}
\label{subsec:sampling}
After assigning $N$ datapoints to $m$ partitions $\mCm$, we move to the next stage where one or more samples are drawn from each partition according to our proposed sampling rule. The primary goal of such a sampling scheme is to explore and discover latent processes by sequentially sampling from the partitions obtained in the previous step. 

\paragraph{Induced graph on the partitions}
\label{subsec:graph}
An essential ingredient of our proposed sampling technique is a graph structure based on the partitions on $\mC$. We define an undirected graph $\mG = (V, E)$ induced by the partitions, where nodes are defined to be $m$ partitions $V = \{\mCm\}$ and there is an undirected edge between $\mCi{i}$ and $\mC_{k_j}$, $k_i \neq k_j$ if their closures share more than one point. In other words $|cl(\mC_{k_i}) \cap cl(\mC_{k_j})| >1$. This criterion means that partitions that share only a corner are not considered neighbors. Note that the edges can entirely be determined during the partition phase for hyperrectangle partitions. 

Other forms of graphs are also possible and sometimes necessary to reflect the known underlying manifold structure of the input space. Our proposed sampling procedure is independent of the graph-generating process and hence allows for more flexibility and adaptability to diverse arrays of applications. 

\paragraph{The algorithm} 
\label{subsec:sampling-algo}
We define importance sampling~\cite{neal2001annealed} alike strategy that leverages the spatial dependence among the input values via the partition-induced graph $\mG$. The basic idea is a sampler that traverses through the partitions via edges $E$ in $\mG$ and successively draws a datapoint from a partition by conditioning on the previously sampled datapoints from neighbor partitions. Such a strategy is motivated primarily  by the intent of untangling convoluted latent processes that the data is arising from, without adding an extra computational burden. We note that since a single datapoint is drawn from each partition, the overall complexity of the sampling depends on $N/m$ and $m$ rather than $N$. 

Without loss of generality, we assume that $\mG$ is connected. If it is not, then the sampling stage can be carried out independently for each connected component. 
The sampling stage is an iterative process. We denote the datapoints in partition $\mCi{i}$ as $\xyi{i}$. The sampler is initialized by randomly sampling a datapoint $\xyistar{1}$ from a partition $\mCi{1}$. (The choice of $k_1$ is discussed later.) Next, the sampler moves to a partition that is an unsampled neighbor of $\mCi{1}$. If there is more than one neighbor to choose from, the sampler moves to an available neighbor according to a criterion similar to the initialization step. The chosen partition is denoted by $\mCi{2}$, and the datapoints $\xyi{2}$ in $\mCi{2}$ are weighted by a symmetric Gaussian kernel denoted by $K(\cdot|y_*^{(1)})$ centered at $y_*^{(1)}$. Let $\fyi{i}$ denote the distribution of $y$ in $\mCi{i}$. Then the weighted sampling distribution $\gyi{i}$ is defined as 
\begin{align}
\label{eq:sampling-distribution}
\gyi{i}(y | \mCi{i}) \, \propto \, & \fyi{i}\big(y | \mCi{i}\big) \times \prod_{j \in \mathcal{A}_i} K\big(y | y_*^{(j)}\big), \;\;  i = 2, \ldots, m,
\end{align}
where $\mathcal{A}_i = I\left(\mN(\mCi{i}) \cap \mVi{i-1} \right)$, 
$\mN(\cdot)$ are the neighbors of $\cdot$ in $\mG$, $\mVi{i}$ is the set of the first $i$ partitions visited by the sampler, $I(\cdot)$ is the index set of $\cdot$, and $K\big(y| y_*^{(j)}\big) \propto \exp \{-\frac{1}{2\eta^2}\big(y - y_*^{(j)}\big)^2 \}$. $\eta$ controls the width of the kernel $K$. Then a sample is drawn from $\mCi{2}$ according to $\gyi{2}$, and the sampler walks through the partitions until all partitions are visited. We note that in practice we set $\eta$ to a value less than the standard deviation of $y_*^{(j)}$ in \eqref{eq:sampling-distribution}.

As previously noted, one of the objectives of such a sampling scheme is to  discover and model latent processes that the data might be arising from. Our initialization criterion is geared toward facilitating this goal. The initial partition is chosen to be one for which the variance of $y$ within the partition is maximal. In other words,
\begin{align*}
\mCi{1} \in \arg \max_{\mC_k \in \mCm} \mbox{Var}(y | \mC_k).    
\end{align*}
Subsequent selection of  partitions is carried out in a similar way. At any iteration, the next sampling partition is chosen from the unvisited neighbors of sampled partitions that have maximum variance,   
\begin{align*}
\mCi{i} \in \arg \max_{\mC_k \in \mN(\mVi{i-1}) \cap \mVi{i-1}^C} \mbox{Var}(y | \mC_k), \;\; i = 2, \ldots, m.
\end{align*}
This process
\ifbool{arxiv}{
, summarized in Algorithm~\ref{algo:sampling}, 
}{}
continues until samples are drawn from all partitions. The weighting strategy of datapoints based on its neighbors encourages the sampling scheme to discover latent global structures in the input-output relationship. 

In a complete setup, this sampling is performed multiple times (independently and in parallel), generating multiple datasets to train our statistical model.
\ifbool{arxiv}{
\begin{algorithm}[t]
    \SetAlgoLined
    \DontPrintSemicolon
    \SetKwInOut{Input}{input}
    \SetKwInOut{Output}{output}
    \SetKwInOut{Init}{initialize}
    \SetKwBlock{Begin}{begin}{end}
    \KwData{$(\xb_1, y_1), \ldots, (\xb_N, y_N)$.}
    \Input{partitions $\mCm$ of $\mC$, \\ 
            associated connected graph $\mG$.}
    \Output{a sample of size $m$, $\xyistar{1}, \ldots, \xyistar{m}$.}
    \Init{
        $\mCi{1} \in \arg \max_{\mC_k \in \mCm} \mbox{Var}(y | \mC_k)$,    \\
        sample $\xyistar{1}$ from datapoints in $\mCi{1}$.
    }
 
 \For{$i = 2,\ldots, m$}{
    \tcp{Select a partition}
    $\mCi{i} \in \arg \max_{\mC_k \in \mN(\mVi{i-1}) \cap \mVi{i-1}^C} \mbox{Var}(y | \mC_k)$, \;
    \tcp{Compute sampling distribution} 
    $\displaystyle \gyi{i}(y | \mCi{i}) \propto \fyi{i}\big(y | \mCi{i}\big) \times \prod_{j \in I(\mN(\mCi{i}) \cap \mVi{i-1})} K\big(y | y_*^{(j)}\big)$, \; 
    \tcp{Draw a sample} 
    $\xyistar{i} \sim \gyi{i}(\cdot)$ from datapoints in $\mCi{i}$.
 }
 \caption{Conditional sampling on graph partitions}
 \label{algo:sampling}
\end{algorithm}
}{}

\subsection{Modeling via ensembles}
\label{subsec:model}
After a set of training data is generated, the last stage is to train a regression model. Our model of choice is Gaussian process~\cite{Rasmussen2005}, which is a semi-parametric regression model, fully characterized by a mean function and a covariance function. Historically GP models have been popular in both ML and statistics because of their ability to fit a large class of response surfaces~\cite{wang2005gaussian,bernardo1998regression,brahim2004gaussian,sacks1989design,gramacy2020surrogates}. The covariance function in a GP model often does the heavy lifting of describing the response variability by means of distance-based correlations among datapoints. Certain classes of GP covariance structures ensure smoothness and continuity in the response surface.

We define $y$ to be a noisy realization from GP $z$. Then the data model can be written as
\begin{align*}
 y(\xb)  = z(\xb) + \epsilon, \;\; \epsilon \overset{iid}{\sim} \mbox{N}(0, \sigma^2)\qquad   
 z(\xb)  \sim  GP(0, \mC_\Phi(\xb, \xb')),
\end{align*}
where $\mC_\Phi(\cdot, \cdot)$ is a covariance function with length-scale parameters $\Phi$. The likelihood of the data is then given by the probability density function of a multivariate normal distribution:
\begin{align}
\label{eq:gplik}
\yb^T = (y_1, \ldots, y_n)^T \sim \mbox{MVN}(\mathbf{0}, \mC_n),
\end{align}
where  $\mC_n$ is an $n\times n$ matrix, obtained by
$
\mC_n = \big[\mC_\Phi(\xb_i, \xb_j) + \sigma^2 \delta_{i=j}\big]_{1 \leq i, j \leq n}.
$
Distribution of $z$ at an untried input setting $\xb^*$ conditioned on $n$ observations is also Gaussian, with mean and covariance given by
\begin{align}
\label{eq:gppred}
    \begin{split}
        \mbox{E}(y(\xb^*) | \xb, \zb, \cdot) & =  c_n(\xb^*)^T \mC_n^{-1} \zb, \\
        \mbox{Var}(y(\xb^*) | \xb, \zb, \cdot) & = \mC_n - c_n(\xb^*)^T \mC_n^{-1} c_n(\xb^*),
    \end{split}
\end{align}
where $c_n(\xb^*) = (\mC_\Phi(\xb^*, \xb_1), \ldots, \mC_\Phi(\xb^*, \xb_n))$. Our implementation uses a scaled separable Gaussian covariance kernel 
$\mC_\Phi(\xb, \xb') = \exp \Bigg( - \sum_{p=1}^d \frac{(x_p - x'_p)^2}{\phi_k} \Bigg). $
The length-scale parameter $\Phi = (\phi_1, \cdots, \phi_d)$ controls the correlation strength along each dimension. Despite a GP's attractive properties, a major drawback of the standard GP model is the associated computational cost for estimating its parameters (i.e., $\Phi$ and $\sigma$). Each evaluation of the likelihood \eqref{eq:gplik} involves inverting $\mC_n$---an operation of $O(n^3)$ complexity. Hence, model training (i.e., estimation of $\Phi$) becomes computationally infeasible as $n$ grows. Alternatives have been proposed to deal with large $n$,  including local GP approximations~\cite{gramacy2015local}, knot-based low-rank representation~\cite{banerjee2012bayesian,snelson2005sparse}, process convolution~\cite{higdon2002space}, and compactly supported sparse covariance matrices~\cite{kaufman2011efficient}. While all these methods achieve a certain computational efficiency at large $n$, none  is intended to discover and model all the latent stochastic processes that the data is possibly arising from. Dirichlet process-based mixture models attempt to solve this problem~\cite{rasmussen2001infinite}, but scalability remains a challenge. In contrast, our proposed method uses $n=m \ll N$ training samples for each GP model, and the sampling scheme actively searches for all latent smooth processes that can be inferred from the data.

\begin{figure}
    \centering
    \includegraphics[trim=150 0 100 0, clip, width = 0.4\linewidth]{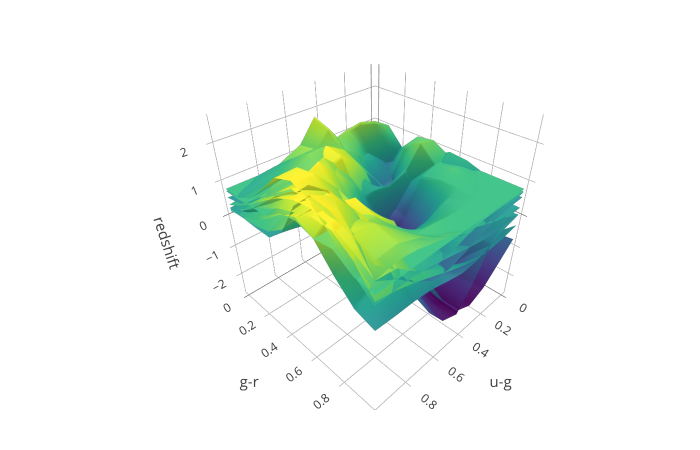}
    \includegraphics[width = 0.55\linewidth]{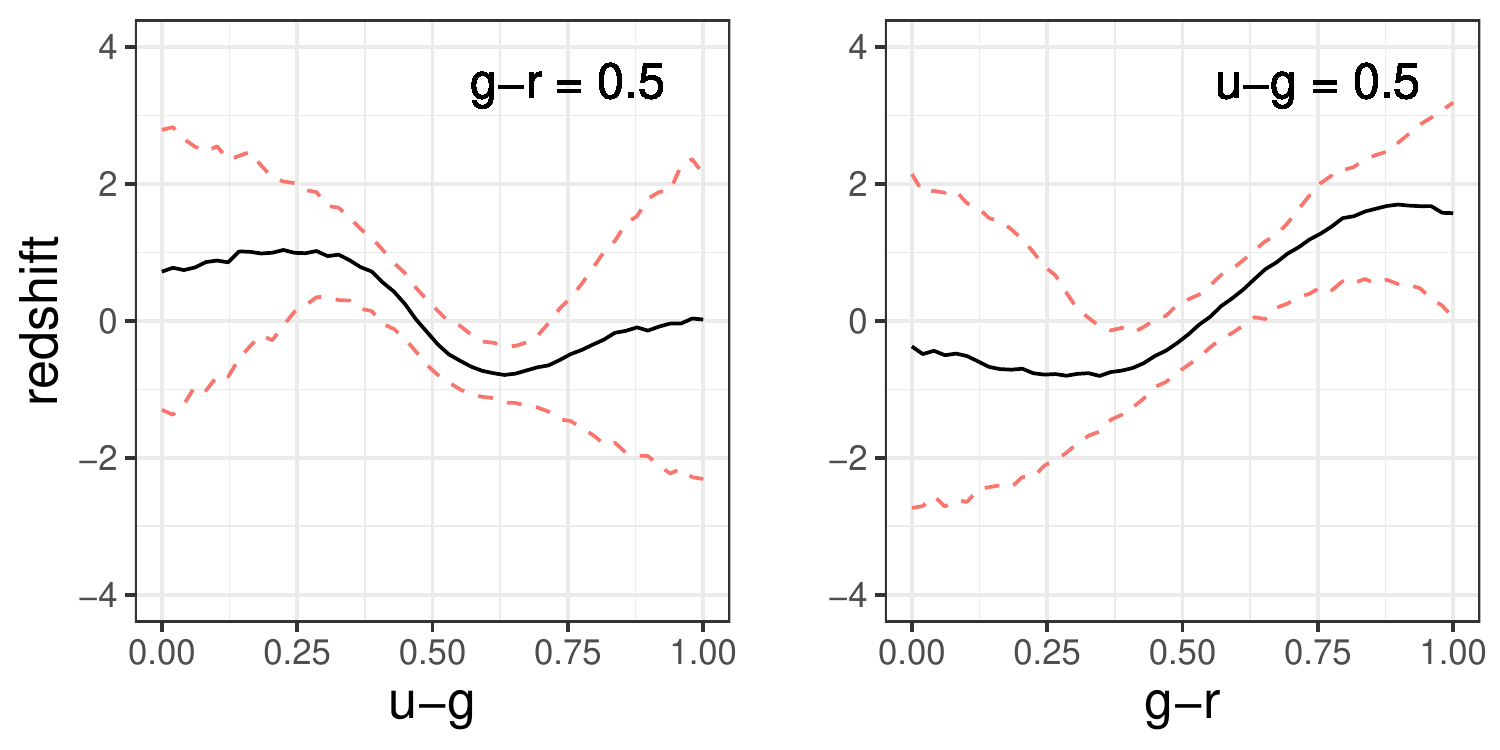}
    \caption{Example ensemble of size 10 of mean GP surfaces;
    inputs are scaled to [0,1]$^2$, and the response is transformed to mean 0 and standard deviation 1. A 1D view of ensemble predictions of redshift at different values of \mbox{$u$-$g$} and $g$-$r$  while keeping the other fixed at 0.5 is shown at the bottom. Black solid lines and red dashed lines represent median prediction and 90\% confidence interval, respectively.}
    \label{fig:bal-kmeans-2d-surf}
\end{figure}

\paragraph{Full predictive model}
Our final predictive model is constructed by taking an ensemble of $N_m$ trained GPs models (each of these models trained with $m$ samples as above). Denoting the predictive distribution at a new input $\xb^*$ as $f_i(y^* | \xb^*)$, we define the ensemble predictor as
\begin{align}
    \hat{f}(y^* | \xb^*) = \frac{1}{N_m} \sum_{i=1}^{N_m} f_i(y^* | \xb^*).
    \label{eq:predictor}
\end{align}
Each $f_i$ has a Gaussian distribution with mean and variance given by \eqref{eq:gppred}. Prediction intervals are computed based on this mixed Gaussian distribution. The individual GPs can be trained independently and hence can be done in parallel. The choice of $N_m$ can be guided by the computing budget and the choices of $\nmin, \nmax$ (which induce a value $m$). Figure~\ref{fig:bal-kmeans-2d-surf} shows an illustration of 10 GP surfaces on a 2D color space, each trained on 50 datapoints sampled from  data of size 2,000. Predicted distribution of redshift at any tuple $(u$-$g, g$-$r)$ is then a combination of Gaussian distributions from the 10 GP models. We note that the estimated prediction uncertainty from ensemble models would be much larger compared with that of a single model on the full data. Moreover, in a single GP case, the family of predictive distributions is  restricted to the family of unimodal symmetric Gaussian distributions, which is often inappropriate for noisy data. In contrast, the ensemble estimate can be interpreted as an approximation of the unknown distribution of the  redshift conditional on the colors in a semi-parametric way.

\section{Results}
\label{sec:results}

In this section we  discuss numerical results from our proposed approach to model the redshift as a function of colors. After removing outliers, the size of the final dataset was reduced to 99,826, of which 20\% were held out for out-of-sample prediction. Predictive distributions for the redshift conditional on colors were constructed by using our implementation of balanced hyperrectangle partitioning and sampling in \texttt{R}~\cite{rcore}; GP models were fitted by using the \emph{mleHomGP} function in the \texttt{hetGP} package~\cite{hetgp}. 

\begin{figure}
\begin{minipage}[t]{.48\textwidth}
    \centering
    \includegraphics[width = \linewidth]{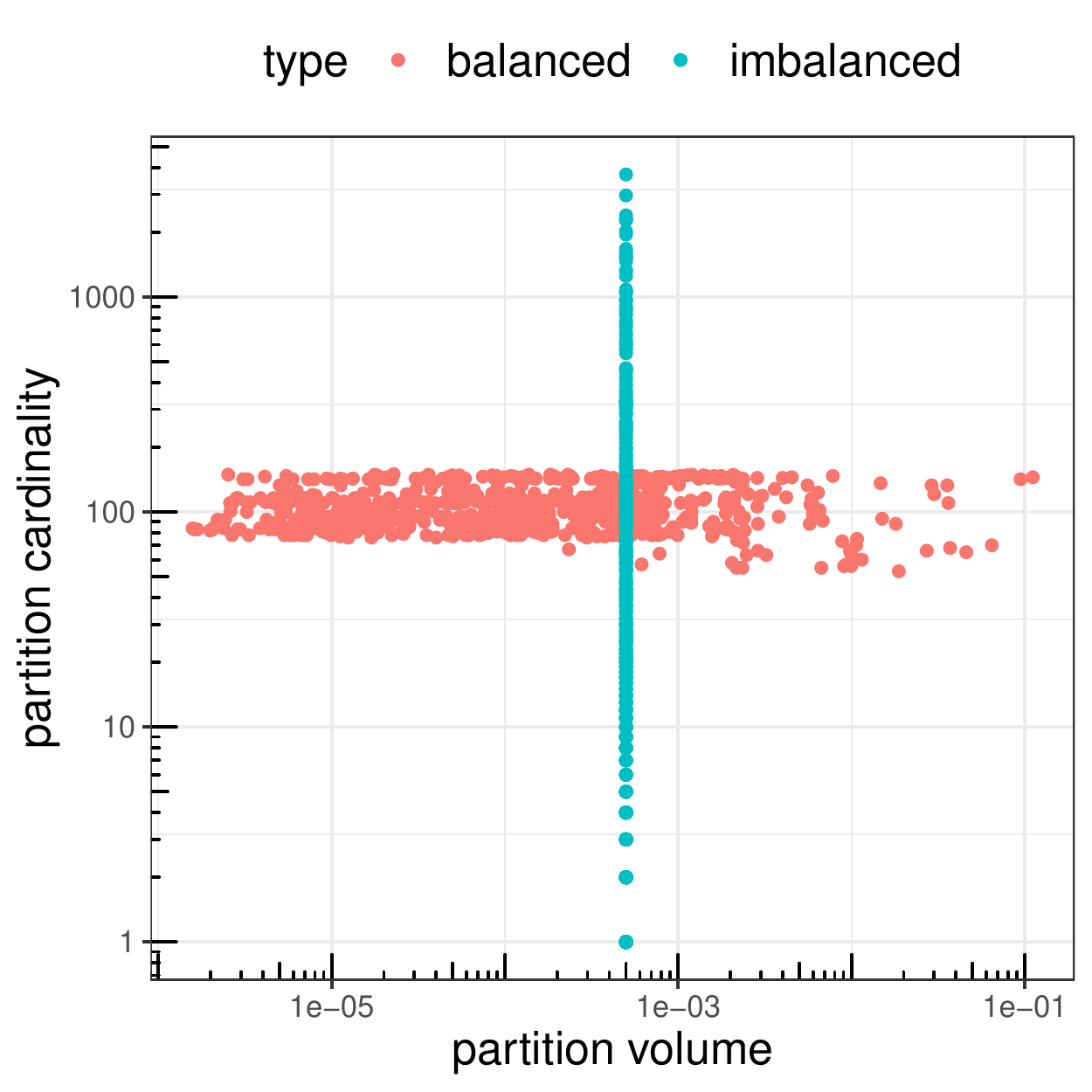}
    \captionof{figure}{
    Properties obtained from 
    balanced hyperrectangle (in red) and equal-volume partitioning (in cyan), with 749 and 711 nonempty partitions, respectively.
    }
    \label{fig:balance-imbalanced-scatter}
\end{minipage}%
\hfill 
\begin{minipage}[t]{.48\textwidth}
  \centering
    \centering
    \includegraphics[width = \linewidth]{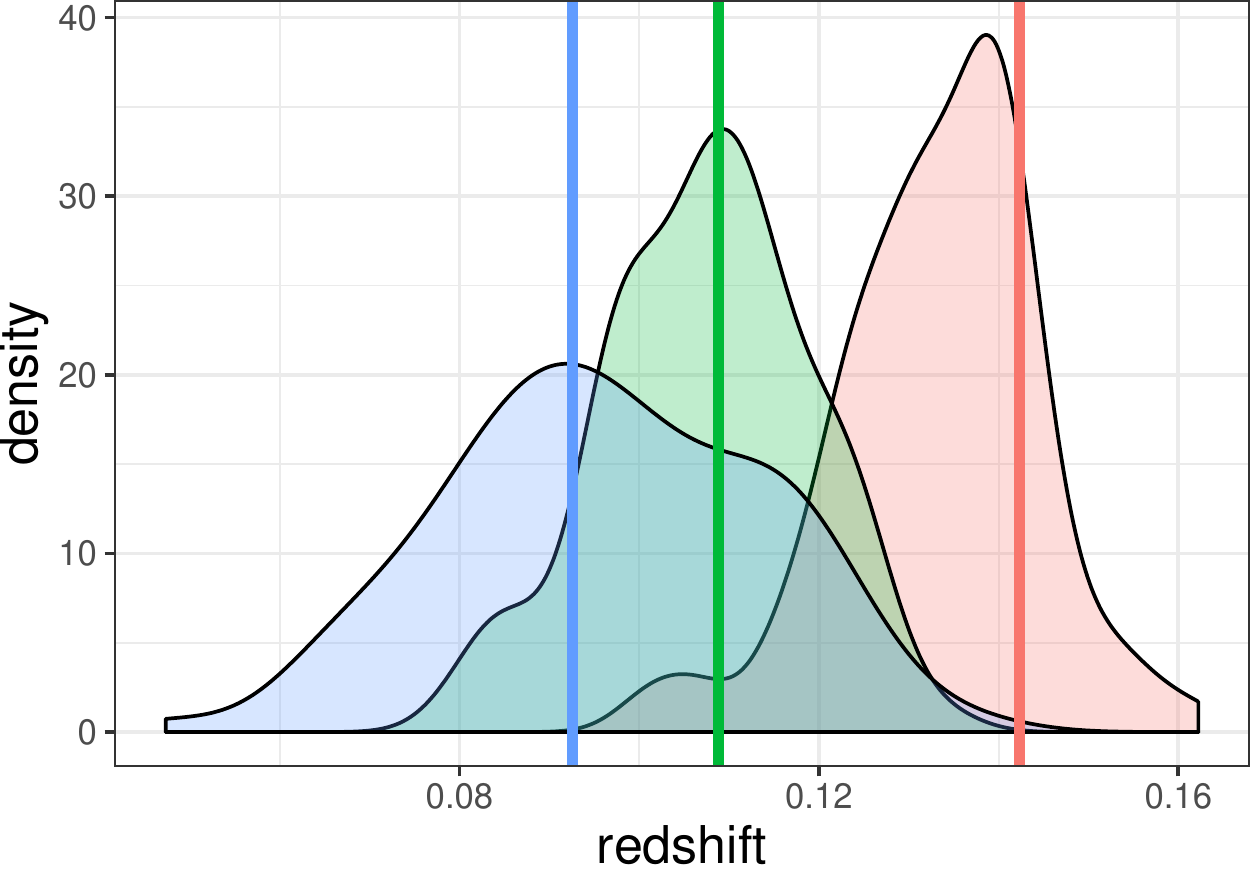}
    \captionof{figure}{Predicted density of redshift at three different color inputs (each indicated by a different shading in the figure). The vertical lines represent the truth for each color input.}
    \label{fig:pred-dens}
\end{minipage}
\end{figure}

\paragraph{Partition and sample}
Inputs to the hyperrectangle partitioning scheme $\nmin$ and $\nmax$ were set at 50 and 150, respectively. These yielded a total of 749 nonempty rectangular partitions with an average cardinality of 107 in the 5D color space. Considering the tradeoff of execution time in the partition step and accuracy of the final prediction, we arrived at this particular choice of $\nmin$ and $\nmax$ after a few trials. At each iteration of sampling and modeling, we drew one sample per partition. The samples  were used to train a zero-mean GP model with a nugget term~\cite{hetgp}. Effectively, each individual GP model was trained on 749 examples. In contrast to partitions being balanced in terms of cardinality, partitions can be constructed to be balanced with respect to volumes, which for hyperrectangles equates to dividing the input space into equal-volume hyperrectangles. The scatter plot in Figure~\ref{fig:balance-imbalanced-scatter} shows the cardinality and volume of partitions as points on $\R^2$ from balanced hyperrectangle and equal-volume partitions. By fixing the number of partitions and their volume, equal-volume partitioning returned 5,769 empty and 711 nonempty partitions with cardinality ranging from 1 to 10,000. Datapoints were not uniformly distributed along any dimension, thus making the equal-volume partitions  have highly imbalanced data compositions. A sample from equal-volume partitions would always result in a biased training set for a model. In contrast, balanced hyperrectangle partitioning optimizes over cardinality by adaptively forming the hyperrectangles that enforce roughly homogeneous data size across partitions.

\begin{figure*}[t!]
    \centering
    \includegraphics[width = 0.48\textwidth]{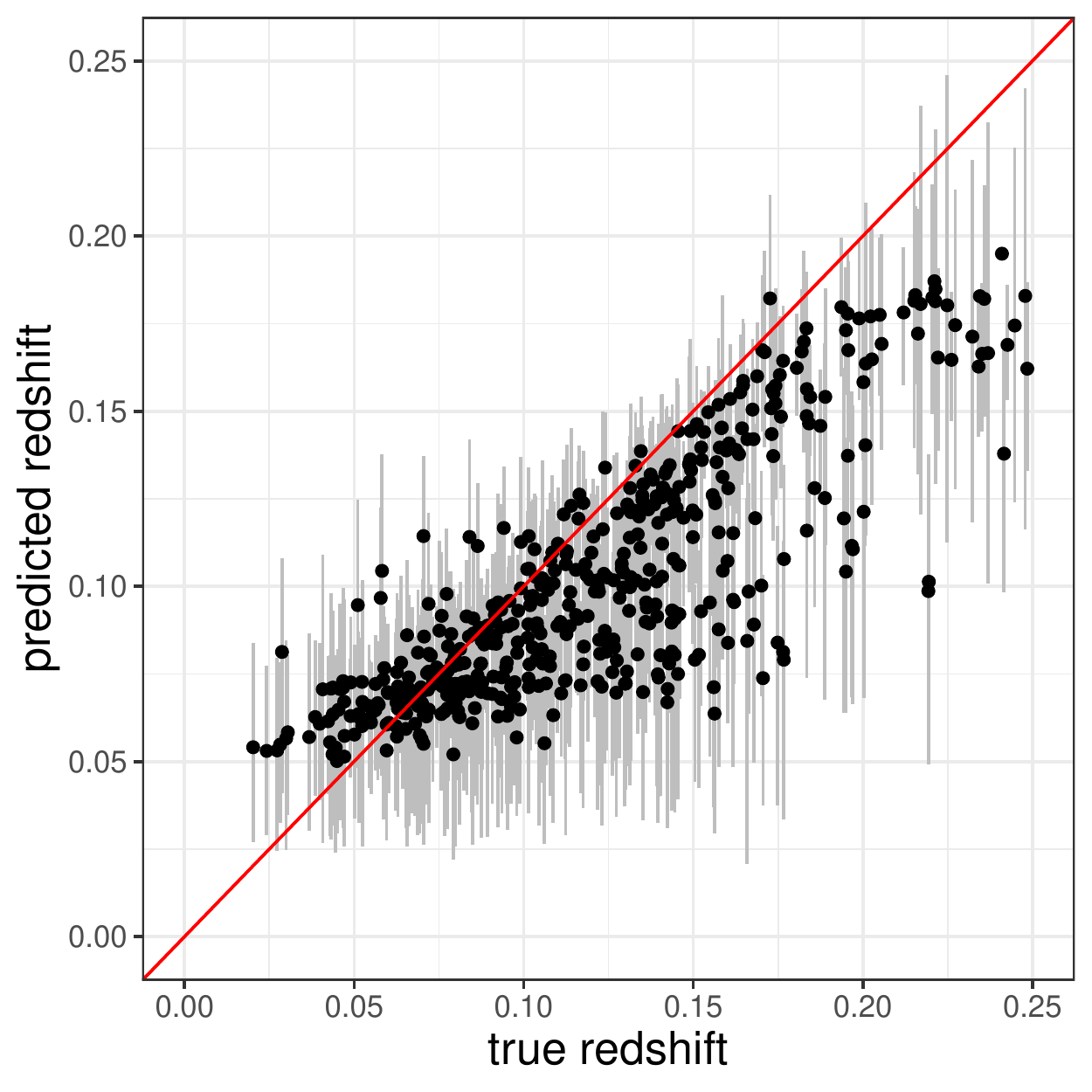} \hfill
    \includegraphics[width = 0.5\textwidth]{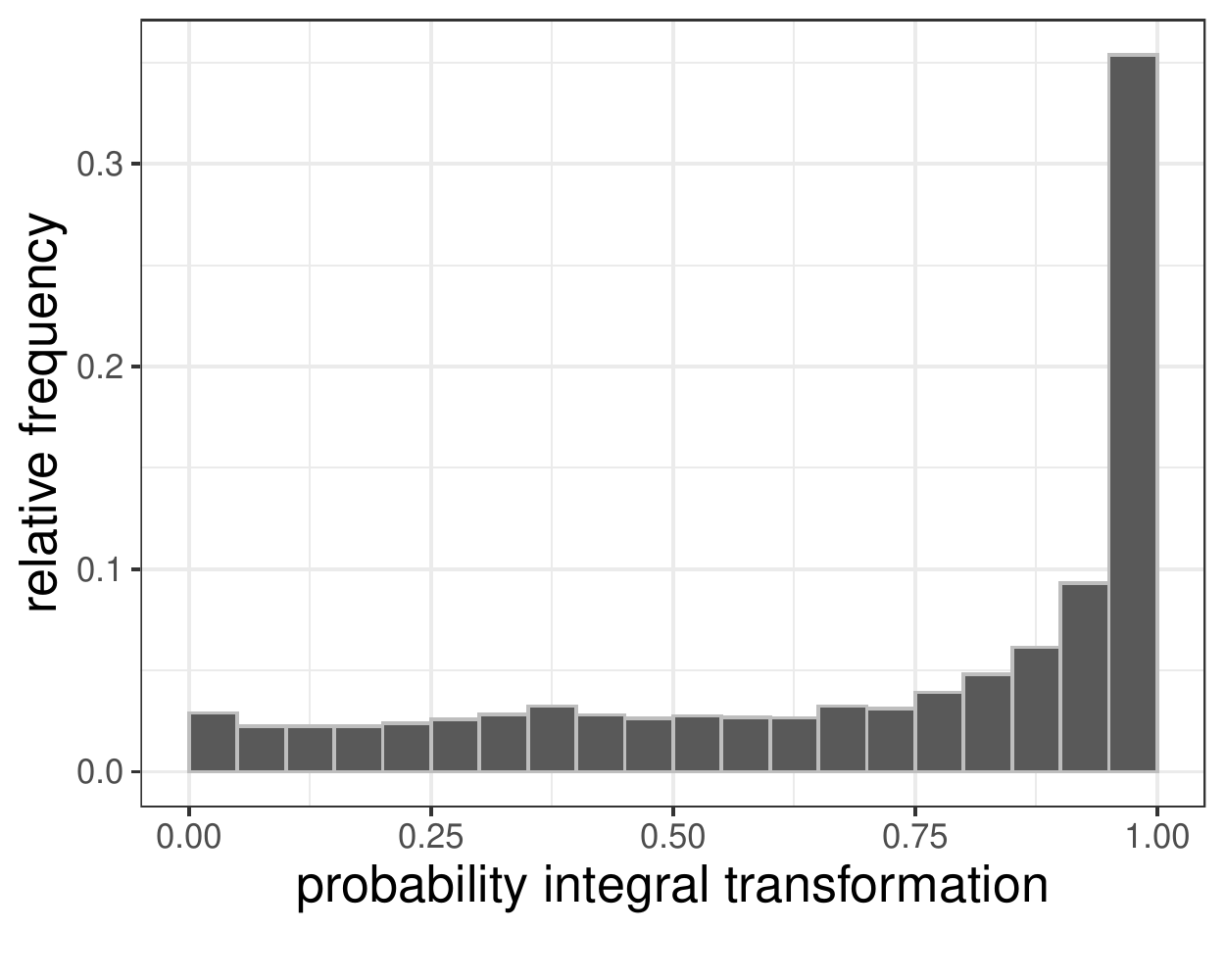}
    \caption{True versus predicted redshifts for 500 random test points. Median predicted redshifts are denoted by black dots, and vertical grey bars show 90\% prediction intervals on the left. A probability integral transformation (PIT) histogram plot is on the right.}
    \label{fig:pred_hyperrec}
\end{figure*}

\paragraph{Model and prediction}
As described in \S~\ref{subsec:model}, zero-mean GP models were fitted to each data subsample for $N_m = 50$ times. Maximum likelihood estimates of length-scale parameters $\Phi$ and the noise variance $\sigma^2$ were obtained and used to construct the ensemble estimate in \eqref{eq:predictor}. Figure~\ref{fig:pred-dens} shows examples of predicted distributions at three different 5D color inputs and the true redshifts. To obtain smooth empirical density estimates, we sampled from the distribution given by \eqref{eq:predictor}; the samples  were then used in a kernel density estimator to construct the predicted density function. As expected, the distributions show asymmetric, multimodal behaviors that would be impossible to capture by a single GP model. On the other hand, unlike in a traditional setting, comparing mean prediction with the true response would not be appropriate here and would result in low prediction accuracy. In other cases, degeneracies in the prediction are resolved by considering the highest mode among all mixture components or the mean from the mixture component with the highest probability~\cite{d2018photometric}. Figure~\ref{fig:pred_hyperrec} shows the prediction performance of 500 random test (out-of-sample) examples. For each example, we obtained the median prediction (denoted by black dots) and 90\% high-density probability region (denoted by vertical grey bars). Because  of the selection effect of observing cosmological objects in a restricted region, imbalance arises with respect to cosmological scale~\cite{d2018photometric}. Objects with higher redshifts are usually less represented in the full dataset in consideration. This is also evident from the probability integral transformation (PIT) plot in Figure~\ref{fig:pred_hyperrec}. The PIT plot is used to assess the quality of the probabilistic predictions---a perfect uniform distribution suggests perfect accuracy. In our case it is far from uniform, because of the shift in prediction at higher redshifts.

\paragraph{Scalability}

Our framework allows for fairly straightforward parallel computations; each stage is amenable to parallelism. Once the data is partitioned according to any partitioning scheme, the ensemble members can be obtained in an embarrassingly parallel way. Each ensemble member consists of sampling data from the partitions and training the GP model, independently from other ensemble members. Almost linear speedup can be achieved when each ensemble member is processed in parallel by using multiple cores in a single computing unit.

\section{Discussion and conclusion}
\label{sec:discussion}

In the SDSS dataset considered here, redshift ranges from 0 to 0.3, with a high concentration around 0.1. To make sense of the Gaussianity assumption for each GP model, we modeled logged redshifts instead of raw ones. Input values were transformed to $[0,1]$ to have consistent length-scale parameter estimates in the covariance. These transformations were handled at the beginning of the data-preprocessing step, before splitting the data into training and testing sets. Such an approach ensured that no extrapolation was being done in predicting the redshifts for the testing examples. Selection of $\nmin$ and $\nmax$ for balanced hyperrectangle partitioning is conceived to be a sequential updating process. In most applications, the choices are highly influenced by the intended size of the training set for each individual model in the ensemble. Prior knowledge of the expected smoothness of the global response surface can assist in choosing the size of the data subsample. For example, estimating a smooth surface using a GP with Gaussian covariance would require a substantially small number of training examples. We used $\nmin = 50$ and $\nmax = 150$ based on prior experience. Construction of the graph induced by partitions is simple and intuitive. One could argue for  considering two hyperrectangles $\mC_{k_i}$ and $\mC_{k_j }$ to be neighbors  when their boundaries intersect only at a corner, namely, $|cl({\mC_{k_i}}) \cap cl({\mC_{k_j}})| = 1$. In fact, neighbor relationships on the partitions can be as arbitrary as an application desires. Considering corner-sharing partitions as neighbors will produce a graph with many more edges, which in turn requires more compute time in the subsequent sampling and modeling stages.

Here we have presented a novel approach to regression modeling for large data that leverages efficient data sampling and an advanced statistical model such as GP modeling. The problem of estimating the full predictive distribution of the photometric redshift provides a base case for demonstrating novel aspects of our proposed methodology. Redshift estimation based on photometric surveys is an important problem. Almost all recent estimation protocols emphasize being able to use large photometry data to train their models~\cite{d2018photometric,beck2016photometric,kaufman2011efficient}: since the new generation of telescopes promises a much larger survey area of the sky, models need to ingest this huge amount of data and produce forecasts for new observations in a reasonable amount of time. One way to meet this need is by using a combination of new algorithms and more powerful supercomputers. 
New modeling paradigms also need to be developed that find a sweet spot between careful choices of efficient algorithms, good models, and opportunity to scale. Indeed, the present work is motivated by this pursuit.

Another aspect of our work involves prediction targets, which are different from simply estimating the mean or median under parametric uncertainty assumptions. Best-case scenarios would involve nonparametric approximation of the unknown predictive distribution, an example of which can be found in Dirichlet process-based models~\cite{gelfand2005bayesian}. Full Bayesian estimation here is rarely feasible with large data. Combining several simple models is a step toward the same goal but at a fraction of the cost. Successful ensemble techniques~\cite{hastie2009multi,liaw2002classification,gramacy2015local} and ML models~\cite{lawrence1997face,dietterich2002ensemble} achieve superior accuracy by combining locally learned structures in the response surface. While we draw motivation from such endeavors, our approach differs from them in a significant way. Each individual model in our ensemble does not target local structures but, rather, learns a global response surface every time a new model is trained on a sample of the data, allowing for a broad range of distribution to be covered. To the best of our knowledge, this work is the first of its kind that draws local inference by combining global information. Evaluating the accuracy of different combinations of partitioning schemes, statistical models, and ensemble sizes is a promising direction for future work.

\subsubsection*{Acknowledgments}
 This material was based upon work supported by the U.S.\ Department of Energy, Office of Science, Office of Advanced Scientific Computing Research, Applied Mathematics and SciDAC programs under Contract No.\ DE-AC02-06CH11357. 
 
\bibliographystyle{splncs04}
\bibliography{reference.bib}

\end{document}